# Global circulation as the main source of cloud activity on Titan

Sébastien Rodriguez[1,2], Stéphane Le Mouélic[1,3], Pascal Rannou[4,5], Gabriel Tobie[1,3], Kevin H. Baines[6], Jason W. Barnes[7], Caitlin A. Griffith[8], Mathieu Hirtzig[9], Karly M. Pitman[6], Christophe Sotin[1,6], Robert H. Brown[8], Bonnie J. Buratti[6], Roger N. Clark[10], Phil D. Nicholson[11]

*[1] Laboratoire de Planétologie et Géodynamique, Université de Nantes, France. [2] Laboratoire AIM, Université Paris 7, CNRS UMR-7158, CEA-Saclay/DSM/IRFU/SAp, France. [3] CNRS, UMR-6112, France. [4] Groupe de Spectrométrie Moléculaire et Atmosphérique, CNRS UMR-6089, Université de Reims Champagne-Ardenne, France. [5] LATMOS, CNRS UMR-7620, Université Versailles-St-Quentin, France. [6] Jet Propulsion Laboratory, California Institute of Technology, Pasadena, CA, USA. [7] NASA Ames Research Center M/S 244-30, Moffett Field, CA 94035. [8] Lunar and Planetary Laboratory, University of Arizona, Tucson, AZ, USA. [9] AOSS, PSL, University of Michigan, Ann Arbor, MI, USA. [10] USGS, Denver Federal Center, Denver, CO, USA. [11] Cornell University, Astronomy Department, Ithaca, NY, USA.*

**Clouds on Titan result from the condensation of methane and ethane and, as on other planets, are primarily structured by the atmosphere circulation[1-4]. At present time, cloud activity mainly occurs in the south (summer) hemisphere, arising near the pole[5-12] and at mid-latitudes[7,8,13-15] from cumulus updrafts triggered by surface heating and/or local methane sources, and at the north (winter) pole[16,17], resulting from the subsidence and condensation of ethane-rich air into the colder troposphere. General Circulation Models[1-3] predict that this distribution should seasonally change moving from an hemisphere to another on a 15-year timescale, and that clouds should develop under certain circumstances at temperate latitudes (~40°) in the winter hemisphere[2]. The models, however, have hitherto been poorly constrained and their long-term predictions have not been observationally verified yet. Here we report that the global spatial cloud coverage on Titan is in general agreement with the models, confirming that cloud activity is mainly controlled by the global circulation. The non-detection of clouds at ~40°N latitude and the persistence of the southern clouds while the southern summer is ending are, however, both in contradiction with models predictions. This suggests that Titan's equator-to-pole thermal contrast is overestimated in the models and that Titan's atmosphere responds to the seasonal forcing with a greater inertia than expected.**

The Visual and Infrared Mapping Spectrometer[18] (VIMS) onboard Cassini provides a unique opportunity to regularly and accurately chart cloud activity from a close vantage point, hence with high spatial resolution and good spectral coverage. We developed a semi-automated algorithm to isolate clouds from other contributions in VIMS images (cf. Fig. 1) and applied it to 10,000 images of Titan. These images encompass several millions of spectra, acquired during 39 monthly flybys of Titan between July 2004 and December 2007.

The total distribution of cloud events derived from our detections (Fig. 2) and the time variation of their latitudinal distribution (Fig. 3a) indicates that cloud activity is clustered at three distinct latitudes during the 2004-2007 period: the south polar region (poleward of 60°S), the north polar region (pole-ward of 50°N), and a narrow belt centered at ~40°S. Individual detection maps are provided for each flyby in the online supplementary information materials (Fig. S1 to S4).

Our study clearly shows the stability of the north polar cloud, which is systematically detected over the 2004-2007 period. We observe this extensive meteorological system poleward of 50-60°N. All of these clouds spectrally differ from the southern clouds, which are presumably formed by wet convec-tion and made of large, tens of microns in size, liquid/solid methane droplets[2,16]. They produce much less signal at 5-µm than any other cloudy features we detect elsewhere on Titan, indicating a lower backscattering at 5-µm. Given that complex indices of refraction of methane and ethane are not that different at this wavelength, the difference in backscattering comes essentially from the particle size. A realtive lower backscattering at 5-µm is consistent with north polar clouds composed of smaller, mi-cron-sized, particles more probably made of solid ethane[2,16,17]. We also detect small elongated clouds at ~60-70°N in March and April 2007. Surrounded by the large north polar ethane cloud, these clouds are thought to be convective methane clouds connected to the underlying lakes[19]. Their higher brightness at 5-µm confirms that they are similar to the methane clouds found in the southern hemisphere.

A few tropical clouds, thought to be rare during Titan's summer, are detected close to the equator (~15°S) on 12 December 2006. Their areas never exceed 10,000 km$^2$. These clouds were therefore un-detectable from ground-based observations. More details about tropical clouds are given in ref. (20). We also observe more than one hundred isolated and transient temperate clouds near 40°S (Fig. 2 and 3a). Most of them are elongated in the east-west direction, as was previously reported[7,8,13-15], possibly due to orographic waves over zonally oriented topography and/or shearing and stretching by strong zonal winds of tens of meter per second[7]. This type of clouds appeared during two periods, in 2004 and then regularly (on the two-thirds of the flybys) between July 2006 and October 2007. Between December 2004 and August 2006, temperate clouds have been observed very rarely (only in October 2005 (ref. 10) and January 2006 (this study)). This could be attributed to the combination of less frequent Titan's flybys by Cassini and/or a momentary decline in cloud activity.

Our latitudinal and time distribution of clouds (Fig.3a) is compared with predictions of the atmospheric Global Circulation Model from ref. (2) (IPSL-TGCM) which is, up to date, the only one to include a microphysical cloud scheme and thus predict the cloud cover (see Fig. 3b). Except for the lack of winter mid-latitude clouds (40°N), we find that the main spatial characteristics of our cloud distribution are well reproduced by the IPSL-TGCM. Clouds appear in the model near 12 km altitude around 40° in the summer hemisphere (the southern hemisphere until 2009), associated to the ascending motion of the convergence zone of a Hadley-type cell[1-3]. Clouds are also predicted very near the summer pole (actual southern) where methane, driven from the warmer region below, condenses generating convective structures[2,21-23]. In the winter polar region, the cloud formation is related to the downwelling stratospheric circulation, which drives an ethane and aerosol enriched stratospheric air into the cold tropopause of the polar night (above 40 km). The observed stability of the north polar clouds is interpreted, with the IPSL-TGCM, as the result of a constant incoming flux of ethane and aerosols from the stratosphere[24], producing a mist of micron-sized droplets of ethane and other products which slowly settles. However, present observations do not confirm the ~40°N clouds predicted by the IPSL-TGCM. In the model, these clouds should result from the horizontal diffusive transport by inertial instabilities of air, partially humid (RH=50%) in tropical regions, toward the colder north pole. At the altitude 12 km, where these clouds are formed, the model predicts $T_{80°N}$-$T_{0°}$ = -4K. Such a contrast makes the air to become saturated and to produce clouds around 40°N. The lack of such clouds in observations could be explained by an actual equator to pole temperature contrast $T_{80°N}$-$T_{0°}$ of about -1.5K instead of the - 4K as predicted by the IPSL-TGCM. Such a small thermal contrast would allow air parcels with RH=50% in tropical regions to move toward the pole without condensing. Conversely, it could also enable the north polar region (where lakes are observed), saturated in methane, to wet the tropical regions up to 50% humidity. If we consider the conditions at the surface, computations, including phase equilibrium with $N_2$-$CH_4$ mixture, show that with an equator-to-pole contrast near the ground of -4.2K (instead of -6.5K in the IPSL-TGCM), an air parcel at methane saturation near the pole (fed by lakes) would be at 50% humidity if transported at tropics. Only 80% humidity would be needed at the north pole if the temperature contrast at surface drops to -3 K, which is actually observed[25].

By contrast, the timing of the summer-hemisphere clouds as constrained by our observations is poorly reproduced by the IPSL-TGCM. Fig. 3b shows that the southern cloud activity should gradually decrease as the equinox approaches, as a consequence of a progressive change in the south polar circulation pattern. This forecasted decline of southern meteorological activity is not supported by our data. According to the IPSL-TGCM, the south polar clouds should have disappeared in mid-2005 and the mid-latitudes clouds should have progressively faded out since 2005, whereas in our observations the southern clouds are still present even late in 2007 and are particularly active at 40°S until mid-2007. The significant latency to the predicted disappearance of summer clouds suggests that the response of Titan's atmosphere to seasonal forcing presents certainly a greater inertia than expected. Yet, since August 2007, south polar clouds' occurrences seemed to be less frequent in our data and the mid-latitude clouds seemed to be scarcer. These very subtle declining trends may indicate that we are witnessing the forthcoming seasonal circulation turnover as we approach the equinox, but with a different timing pattern than forecasted by the IPSL-TGCM.

Fig. 4 shows that, between July 2004 and December 2007, the mid-latitude clouds are not uniformly distributed in longitude, as already noticed during previous ground-based observations[14] (December 2003-February 2005). The clouds' propensity for 0° longitude found in 2003-2005 was attributed to localized geological forcings from the surface possibly related to an active cryovolcanic province[14]. Yet, three years later, our distribution differs markedly, showing more structures (Fig. 4c). Contrary to ref. (14), we observe mid-latitudes clouds at almost all longitudes with an excess at longitudes (from 60°E to 180°E corresponding to the leading hemisphere of Titan) where ref. (14) detected none. The strong clouds' density peak, along with the secondary bump, both reported by ref. (14) have drifted eastward by 30° with an estimated rate of ~10° by terrestrial year. In addition, we found two troughs at longitudes facing Saturn (0°) and anti-Saturn (180°). Though the strong link of the clouds to the latitude indicates that global circulation plays a major role in their formation[1-3], the wavy pattern of our clouds' distribution suggests a secondary forcing mechanism. The 30° longitude shift in the cloud distribution between the periods 2003-2004 (ref. 14) and 2005-2007 (this study), as well as the loose correlation of clouds with surface location, exclude surface geological activity as the primary triggering mechanism. Both the drift in longitude and the discovery of two diametrically opposite minima rather favour processes taking place in Titan's atmosphere, that we attribute to external forcing by Saturn's tides. Saturn's tides are predicted to generate tidal winds in Titan's dense atmosphere, particularly significant in the troposphere[26] at altitudes where temperate clouds are found to develop[2,3,13-15]. These winds manifest themselves as eastward travelling planetary-scale waves of degree-two and change east-west direction periodically through the tidally locked orbit of Titan[26]. In consequence, tidally-induced winds periodically modify the convergence of air masses, mostly at two preferential longitudes 180° apart, potentially resulting in perturbations to cloud formations[26].

The extension of the Cassini mission possibly up to the summer solstice in 2017 and the continuation of ground-based observations will feed the GCMs with further observational constraints. The refined GCMs will provide a better knowledge of the global atmospheric circulation, which is crucial for understanding the carbon-cycle on Titan.

**Supplementary Information** is linked to the online version of the paper at www.nature.com/nature.

**Acknowledgments** We thank M.E. Brown for fruitful discussions that allowed us to greatly improve the quality of this study. This work was partly performed at the Jet Propulsion Laboratory, California Institute of Technology, under contract to the National Aeronautics and Space Administration. KMP and JWB are supported by the NASA Postdoctoral Program, administered by Oak Ridge Associated Universities. Calibrated VIMS data appear courtesy of the VIMS team. We thank the CNRS, CEA and CNES French agencies, as well as the University Paris Diderot for their financial support.

**Author Information** Reprints and permissions information is available at npg.nature.com/reprinsandpermissions. The authors declare no competing financial interests. Correspondence and requests for materials should be addressed to S.R. (sebastien.rodriguez@cea.fr).

## Figure captions:

Figure 1: **Method of spectral detection of Titan's clouds illustrated on a representative VIMS data cube.** The VIMS onboard Cassini acquires a 352-channels spectrum from 0.3 to 5.1 µm for each pixel of an image[18]. (**a**) shows a scatter plot of the 2.75 µm window integrated area versus the 5 µm window integrated area of the VIMS color-image shown in (**b**) with Red=2.03-µm, Green=2.78-µm, Blue=5-µm. The integrated window areas correspond to the integral of I/F within the spectral range shown in gray in spectra. (**c**) and (**d**) correspond to the 2.75-µm and 5-µm integrated window area images respectively, coded in grayscale (high values appear in bright). Characteristic spectra are inseted within (**a**), showing clouds (red), limb (violet), typical surface (cyan) and a high 5-μm signal surface feature (Tui Regio[27]) in green. ”Surface” windows correspond to peaks at 1.27, 1.59, 2.03, 2.75 and 5 μm. Because clouds are efficient reflectors and reduce the path-length of solar photons, their spectra present a brightening of all “surface” windows relative to other spectra. We found that the most robust spectral criterion to separate clouds' pixels from other contributions (surface and limb) is the simultaneous increased integrated areas of the 2.75-μm and 5-μm windows. Conservative, two-sigma thresholds on the integrated areas of these two windows are automatically calculated in order to isolate pixels corresponding to clouds (red triangles in (**a**)). We deliberately choose a conservative threshold to avoid false positives. This can lead to the rare non-detection of optically thin or low-altitude clouds, of clouds much smaller than a VIMS pixel, or of clouds that are too close to the limb. (**e**) shows the resulting cloud pixels detection (in red) which are then reprojected on a global map (see Fig. 2).

Figure 2: **Maps of Titan's clouds derived from VIMS observations from July 2004 to December 2007.** Our detections are presented in cylindrical (top) and polar orthographic (bottom) projections. The colors of the clouds correspond to the date of each cloud observation. A VIMS grayscale mosaic of Titan's surface (adapted from RGB color composite global mosaics in ref. (28)) is used as background. Clouds are found to be distributed in three clustered regions: the two poles and the southern temperate latitudes. Only very few occurrences of clouds are found in equatorial regions. One cloud event is found on December 2005 just above a particularly interesting terrain thought to be of cryovolcanic origin (Tui Regio[27]) and may witness possible recent cryovolcanic activity.

Figure 3: **Latitudinal Titan's cloud coverage with time compared with Global Circulation Model[3] predictions.** ***Top*****:** We reported here the latitudinal distribution of clouds we detected with VIMS versus time from July 2004 to December 2007. The thin blue vertical lines mark the time of the VIMS observations. The latitude extent of the clouds we detect is enhanced with thicker vertical lines, in blue

when in dayside and in green when in polar night. Isolated temperate clouds are colored in purple. The previous Cassini and ground-based observations reported in the literature are superimposed over our latitudinal distribution by colored dots and diamonds respectively. Our detections are in very good agreement with the previous observations. ***Bottom:*** Integrated Titan's cloud opacity above 10 km, summed each year, predicted by ref. (2)'s GCM (IPSL-TGCM) between 2004 and 2011. The thick black lines show the edge of the polar night. Spatial distribution of clouds forecasted by the IPSL-TGCM, confining clouds at the two poles and around 40°S, is in very good agreement with our observations (see ***top*** and Fig. 2). On the contrary, the observed clouds timing is poorly reproduced by the IPSL-TGCM. In the time interval monitored by VIMS for this work, the IPSL-TGCM predict that the south pole cloud should vanish before the equinox for more than one year, and that the 40°S cloud belt should have reached a maximum of intensity between 2004 and 2007 and then should gradually vanish with the incoming circulation turnover. This seems to be lately observed by VIMS, with a significant delay (see text for details).

Figure 4: **The southern temperate clouds distribution in longitudes.** ***(a)****:* The total number observations that cover each 10° bin of longitude is shown with the solid red line for our study and the black dotted line for ref. (14). ***(b)****:* The number of clouds observed by VIMS between July 2004 and December 2007 (our study - solid red line) and ref. (14) between December 2003 and February 2005 (black dotted line) in each 10° bin of planetocentric longitude summed within 60°S and 0° of latitudes. Blue bars indicate the Poisson standard deviation for each VIMS clouds count. The statistics indicate that the overall shape of the longitudinal distribution is significant. ***(c)****:* Normalized numbers of clouds (number of clouds divided by the number of observations) from ref. (14) and from this study are compared. Our distribution shows two minima at the sub- (0°E), where ref. (14) saw a maximum, and anti-Saturn points (180°E). Two others minima are also present in the neighbourhood of 70°E and -110°E longitude. But, due to Cassini's Saturn tour limitation, the detection of clouds was heavily precluded

here by particularly low spatial resolution (Fig. S5a) and very unfavourable conditions of observations (resulting to high airmass – Fig. S5b), so that these two minima cannot be interpreted with confidence.

**Figure 1**

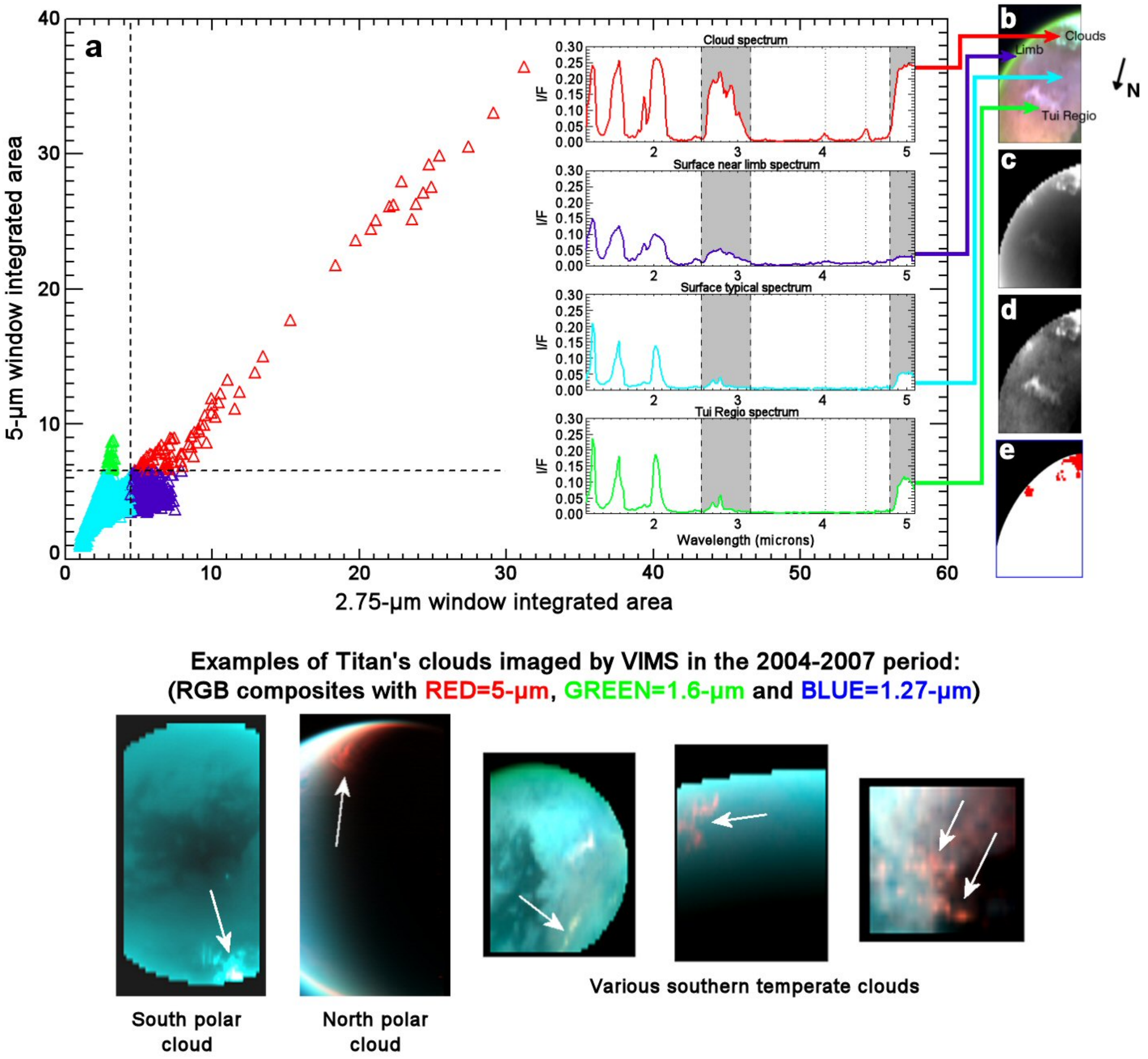
a
b
c
d
e
5-µm window integrated area
2.75-µm window integrated area
Cloud spectrum
Surface near limb spectrum
Surface typical spectrum
Tui Regio spectrum
I/F
Wavelength (microns)
Clouds
Limb
Tui Regio
N
Examples of Titan's clouds imaged by VIMS in the 2004-2007 period:
(RGB composites with RED=5-µm, GREEN=1.6-µm and BLUE=1.27-µm)
South polar cloud
North polar cloud
Various southern temperate clouds

**Figure 2**

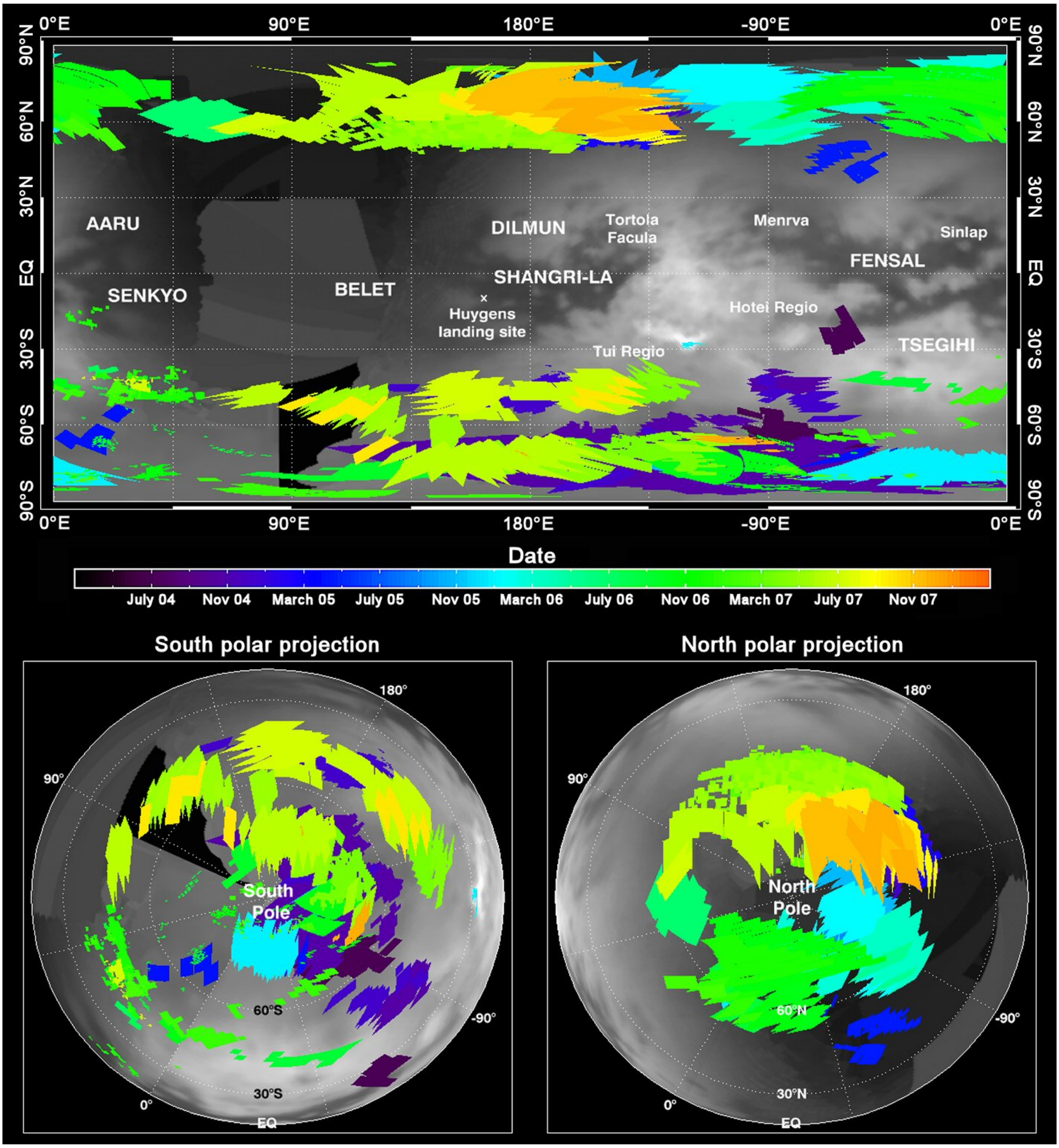

**Figure 3**

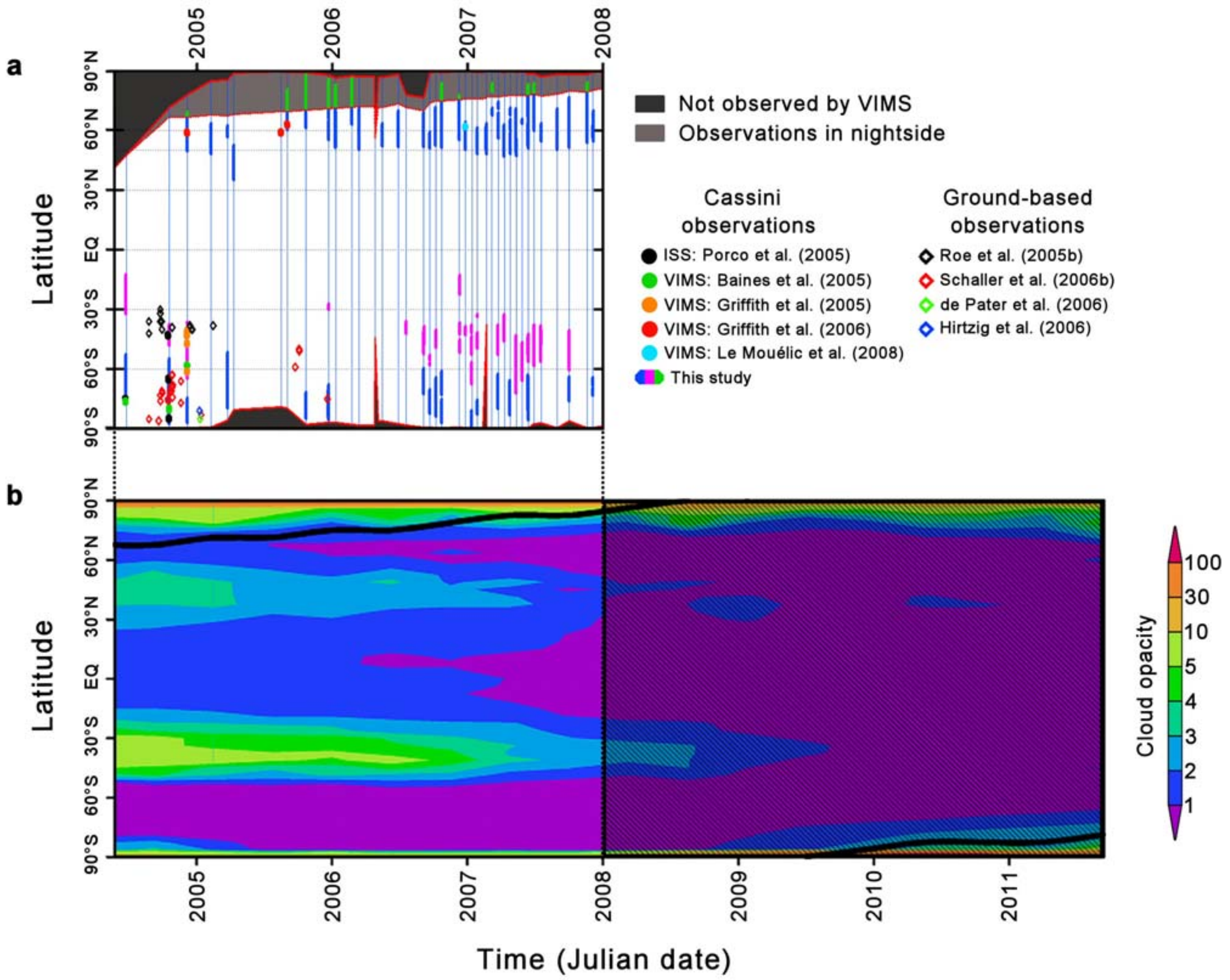
a
Latitude
90°N
60°N
30°N
EQ
30°S
60°S
90°S
2005
2006
2007
2008
Not observed by VIMS
Observations in nightside
Cassini observations
ISS: Porco et al. (2005)
VIMS: Baines et al. (2005)
VIMS: Griffith et al. (2005)
VIMS: Griffith et al. (2006)
VIMS: Le Mouélic et al. (2008)
This study
Ground-based observations
Roe et al. (2005b)
Schaller et al. (2006b)
de Pater et al. (2006)
Hirtzig et al. (2006)
b
2009
2010
2011
Cloud opacity
100
30
10
5
4
3
2
1
Time (Julian date)

**Figure 4**

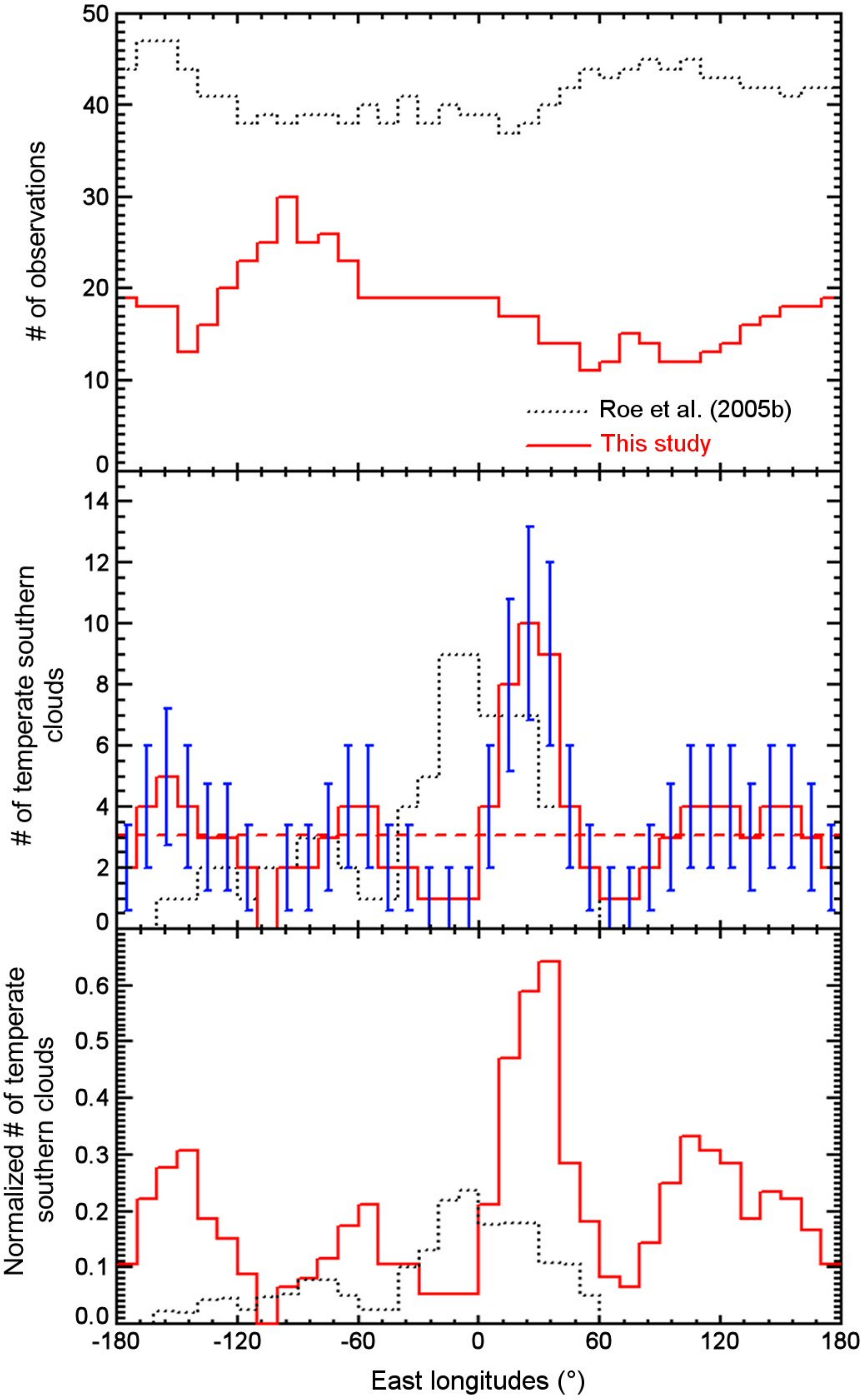

# of observations
Roe et al. (2005b)
This study
# of temperate southern clouds
Normalized # of temperate southern clouds
East longitudes (°)
-180
-120
-60
0
60
120
180